\documentstyle[preprint,prb,eqsecnum,aps]{revtex}
\begin{document}
\tighten
\draft

\title{Spontaneous Interlayer Coherence in Double-Layer
Quantum Hall Systems: Symmetry Breaking Interactions,
In-Plane Fields and Phase Solitons}

\author{Kun~Yang\cite{KY},K.~Moon\cite{KM},
Lotfi~Belkhir\cite{LB}, H.~Mori\cite{HM},S.M.~Girvin,
A.H.~MacDonald}
\address{Department of Physics, Indiana University, Bloomington,
IN~~47405}

\author{L.~Zheng\cite{LZ}}
\address{Department of Physics and Astronomy,
University of Kentucky, Lexington,
KY~~40506}

\author{D.~Yoshioka}
\address{Institute of Physics, College of Arts and Sciences, University of
Tokyo, Komaba, Meguroku Tokyo 153, Japan}

\date{\today}

\maketitle

\begin{abstract}
At strong magnetic fields double-layer two-dimensional-electron-gas
systems can form an unusual broken symmetry state with spontaneous
inter-layer phase coherence.  The system can be mapped to
an equivalent system of pseudospin $1/2$ particles with pseudospin-dependent
interactions and easy-plane magnetic order.  
In this paper we discuss
how the  presence of a weak interlayer tunneling term alters the
properties of double-layer systems when the broken symmetry is
present.  We use the energy functional and equations of motion derived
earlier to evaluate the zero-temperature response functions
of the double-layer system and use our results to
discuss analogies between this system and Josephson-coupled
superconducting films.  We also present
a qualitative picture of the low-energy charged
excitations of this system.  
We show that parallel fields induce a
highly collective phase transition to an incommensurate
state with broken translational symmetry.
\end{abstract}


\section{Introduction}
\label{sec:intro}

The study of correlated electron systems in fewer than three dimensions
continues to be an important theme in condensed matter physics.
In particular, the study of strongly correlated two-dimensional (2D)
electron systems on a lattice has been motivated by high-temperature
superconductivity, while the study
of strong correlations in continuum two-dimensional systems
has been motivated by the fractional quantum Hall effect\cite{fqhe}.
Properties of high-temperature superconductors are thought by some
to be strongly influenced by the weak coupling which exists
between the superconducting planes.  In the fractional
quantum Hall effect, early work by Halperin\cite{halperin} anticipated
novel fractional quantum Hall effects due to interlayer
correlations\cite{gsnum} in multi-layer systems.
Recent technological progress has made it possible to
produce double-layer two dimensional electron gas systems
of extremely high mobility in which these effects can be observed.
The two electron layers can either be bound in
separate quantum wells\cite{murphyPRL} as illustrated schematically
in Fig.~(\ref{fig:fig1}) or bound to opposite edges of a
single wide quantum well.\cite{mansour}
In both cases the 2D electron gases are separated by a distance $d$
small enough ($d\sim 100$\hbox{\AA}) to be comparable
to the typical spacing between electrons in the same layer.

In a large magnetic field, strong Coulomb correlations between the layers
have long\cite{halperin} been expected to lead to novel fractional quantum
Hall effects.  Correlations are especially important in the strong
magnetic field regime because all electrons can be accommodated within the
lowest Landau level and execute cyclotron orbits with a degenerate kinetic
energy. The fractional quantum Hall effect occurs when the system has a
gap for making charged excitations, {\it i.e.} when the system is
incompressible.   Theory has predicted\cite{halperin,gsnum,amdreview} that
at some Landau level filling factors, gaps occur in double-layer systems
only if interlayer interactions are sufficiently strong. These theoretical
predictions have been confirmed experimentally\cite{expamd}. More
recently, theoretical work from several different points of
view\cite{wenandzee,ezawa,ahmz1,gapless,harfok,usPRL,Ilong,jasonho,lopez}
has suggested that inter-layer correlations can also lead to unusual
broken symmetry
states with spontaneous phase coherence between layers which
are isolated from each other (except for inter-layer Coulomb interactions).
We have argued\cite{usPRL} that it is spontaneous interlayer
phase coherence which is responsible for the recently
discovered\cite{murphyPRL} extreme sensitivity of the
fractional quantum Hall effect at total Landau level filling
factor $\nu = 1$ to small tilts of the magnetic field away from
the normal to the layers. ($\nu \equiv N / N_{\phi}$ where $N$ is the
number of electrons and $ N_{\phi}$ is the number of single-particle
levels per Landau level.)
In a previous lengthy paper\cite{Ilong} (hereafter referred to as I)
we have developed a rather complete
description of the novel physics associated with spontaneous inter-layer phase
coherence in the case where there is no tunneling between the layers.  The
present companion paper will analyze the case of finite tunneling between the
layers and the role of magnetic field tilt.  Many of the ideas presented
in detail here are discussed qualitatively in Ref.[\onlinecite{bookchap}].

\section{Experimental Background}
\label{sec:expback}
In this section we review the experimental\cite{jpebook}
 indications that the system
is spontaneously
ordered and exhibits excitations which are highly collective in
nature.  We focus here and throughout this paper on the case of
Landau level filling factor $\nu = 1$ (that is, 1/2 in each layer)
and assume for simplicity that
the electronic spin degrees of freedom are frozen out by the Zeeman
energy.  The schematic energy level diagram for the growth-direction
degree-of-freedom in the double-layer system is shown
in Fig.~(\ref{fig:fig1}) for the case of non-interacting electrons.
For simplicity we assume that electrons
can occupy only the lowest electric subband in each quantum well.
If the barrier between the wells is not too strong,
tunneling from one side to the other is allowed. The lowest energy
eigenstates split into symmetric and antisymmetric combinations separated
by an energy gap $\Delta_{\text SAS}$ which can, depending on the sample,
vary from essentially zero to many hundreds of Kelvins.  The splitting
can therefore be much less than or greater than the interlayer interaction
energy scale, $E_{\text c} \equiv e^2/\epsilon d$.

When the layers are widely separated, there will be no correlations between
them and we expect no dissipationless quantum Hall state, since each
layer has\cite{nuhalf} $\nu = 1/2$.  For smaller separations, it is
observed experimentally that there is an excitation gap and a quantized
Hall plateau.\cite{greg,mansour,murphyPRL}  This has either a trivial or a
highly non-trivial explanation, depending on the ratio
$\Delta_{\text SAS}/ E_{\text c}$.   For large $\Delta_{\text SAS}$
the electrons tunnel back and forth so rapidly that it is as if there is
only a single quantum well.  The tunnel splitting $\Delta_{\text SAS}$
is then analogous to the electric subband splitting in a (wide) single
well.  All symmetric states are occupied and all antisymmetric states are
empty and we simply have the ordinary $\nu = 1$ integer Hall effect.
Correlations are irrelevant in this limit and the excitation gap is close
to the single-particle gap $\Delta_{\text SAS}$  (or $\hbar\omega_{\text c}$,
whichever is smaller).
What is highly non-trivial about this system is the fact that the
$\nu = 1$ quantum Hall plateau survives even when the
tunnel splitting becomes arbitrarily small: $\Delta_{\text SAS}
\ll E_{\text c}$.  In this limit the excitation gap has clearly changed to
become highly collective in nature since the
observed\cite{greg,mansour,murphyPRL}
gap can be on the scale of 20K even when $\Delta_{\text SAS} \sim
1{\text K}$.  As we will see below, because of a spontaneous broken
symmetry,\cite{wenandzee,ezawa,harfok,usPRL} the excitation gap actually
survives the limit $\Delta_{\text SAS} \longrightarrow 0$ as illustrated
in Fig.~(\ref{fig:qhe/no-qhe}).
This cross-over from single-particle to collective gap is,
as we will show, quite analogous to the result that for spin polarized
single layers, the excitation gap survives the limit of zero Land\'e
g factor and hence `$\nu=1$ is a fraction too.'\cite{sondhi}

A second indication of the highly collective nature of the excitations
can be seen in the Arrhenius plots showing thermally activated
dissipation.\cite{murphyPRL}  The low temperature activation energy
$\Delta$ is, as already noted, much larger than $\Delta_{\text SAS}$.  If
$\Delta$ were nevertheless somehow a single-particle gap, one would expect
the Arrhenius law to be valid up to temperatures of order $\Delta$.
Instead one observes a rather abrupt leveling off in the dissipation as the
temperature increases past values as low as $\sim 0.1 \Delta$.
This effect is observed both in double-well systems and wide single-well
systems and is consistent with the notion of a thermally induced
collapse of the order that had been producing the collective gap.

The third significant feature of the experimental data pointing to a
highly-ordered collective state is the strong response of the system
to relatively weak magnetic fields $B_\parallel$ applied in the plane of
the 2D electron gases.  Within a model that neglects higher electric
subbands, we can treat the electron gases as strictly two-dimensional.
(There is ample evidence\cite{nickila} that parallel field effects due to
subband
mixing within a single quantum well produce only small, albeit measurable,
effects.)  $B_\parallel$ can then
affect the system only if there are processes involving tunneling
that carry electrons around closed loops containing flux.  A
prototypical such process is illustrated in Fig.~(\ref{fig:fig2}).  An electron
tunnels from one layer to the other at point A, and travels to point B.
Then it (or another indistinguishable electron) tunnels back and returns
to the starting point.  The parallel field contributes to the quantum
amplitude for this process (in the 2D gas limit) a gauge-invariant
Aharonov-Bohm phase factor $\exp\left(2\pi i \Phi/\Phi_0\right)$ where
$\Phi$ is the enclosed flux and $\Phi_0$ is the quantum of flux.
Such loop paths evidently contribute significantly to correlations in
the system since the activation energy gap is observed to decrease very
rapidly with $B_\parallel$, falling by factors of order two to ten
(depending on the sample) until
 a critical field, $B^*_\parallel \sim 0.8{\text T}$, is
reached at which point the gap
essentially ceases changing.  To understand how remarkably small
$B^*_\parallel$ is, consider the following.  We can define a length
$\xi_\parallel$ from the size of the loop needed to enclose one quantum of
flux: $\xi_\parallel  B^*_\parallel d = \Phi_0$.
($\xi_\parallel [\AA] = 4.137 \times 10^5 / d [\AA] B^*_\parallel
[\text T]$.) For $B^*_\parallel =
0.8{\text T}$ and $d = 150 \AA$, $\xi_\parallel = 0.27 \mu{\rm m}$ which is
approximately twenty times the spacing between electrons in a given layer
and thirty times larger than the quantized cyclotron orbit radius
$\ell \equiv (\hbar c / e B_\perp)^{1/2}$ within an individual layer.
Significant drops in the excitation gap are already seen at fields of 0.1T
implying enormous phase coherent correlation lengths must exist.  Again
this shows the highly collective nature of the ordering in this system
associated with spontaneous interlayer phase coherence.

\section{Spontaneous Phase Coherence and Symmetry-Breaking by Tunneling}
\label{sec:spontaneous}

The essential physics of spontaneous inter-layer phase coherence can
be addressed either from a microscopic point of
view\cite{ahmz1,gapless,harfok,usPRL,Ilong} or from a macroscopic Chern-Simons
field theory point of
view,\cite{wenandzee,ezawa,Ilong}  but it is perhaps most easily explained in
terms of the simple variational wave function
\begin{equation}
|\psi\rangle = \prod_X\left\{c^\dagger_{X\uparrow} + 
e^{i\varphi} c^\dagger_{X\downarrow} \right\} |0\rangle,
\label{eq:variational}
\end{equation}
where $X$ is a state label (for instance,
the Landau gauge orbital guiding center\cite{Ilong})
and we are using a pseudospin notation in which spin up refers
to electrons in the upper layer and spin down refers to electrons in the
lower layer.\cite{Ilong}  The interpretation of this wave function is that
every Landau orbital $X$ is occupied (hence $\nu=1$), but the
system is in a coherent linear combination of pseudospin up and down
states determined by the phase angle $\varphi$.  This means that the
system has a definite total number of particles ($\nu=1$ exactly)
but an indefinite number of particles in each layer.  In the absence of
inter-layer tunneling, the particle number in each layer is a good
quantum number.  Hence this state has a spontaneously
broken symmetry\cite{gapless,wenandzee,ezawa,Ilong}
in the same sense that the BCS state for a superconductor
has indefinite (total) particle number but a definite phase relationship
between states of different particle number.

In the absence of tunneling, the energy can not depend on the phase angle
$\varphi$ and the system exhibits a global $U(1)$ symmetry associated with
conservation of particle number in each layer.  One can imagine allowing
$\varphi$ to vary slowly with position to produce excited states.
Because of the $U(1)$ symmetry, the effective Hartree-Fock
energy functional for these states has a gradient expansion 
whose leading term must have the form
\begin{equation}
H = \frac{1}{2}\rho_s\int d^2r |\nabla\varphi|^2 + \ldots\,\,.
\label{eq:xymod}
\end{equation}
The origin of the finite `spin stiffness' $\rho_s$ is the loss of
exchange energy which occurs when $\varphi$ varies with position.
Imagine that two particles approach each other.  They are in a linear
superposition of states in  each of the layers (even though there is no
tunneling!). If they are characterized by the same phase $\varphi$, then
the wave function is symmetric under pseudospin exchange and so the
spatial wave function is antisymmetric and must vanish as the particles
approach each other. {\it States with spontaneous phase coherence have
better interlayer  electronic correlations and hence lower interlayer
Coulomb interaction energy.}  If a phase gradient exists then there is a
larger amplitude for the particles in opposite layers  to be near each
other and hence the interlayer interaction energy is
higher.\cite{smearingarg} This loss of exchange energy is the source of
the finite spin stiffness and is what causes the system to spontaneously
`magnetize'.

A skeptical reader might legitimately worry that the single-Slater-determinant
variational wavefunctions in terms of which we have framed the
above discussion misrepresent the physics.  Indeed, the ground state 
of double-layer systems in the Hartree-Fock approximation has
spontaneous phase coherence at all values of $\nu_T$ whereas we
believe that this property actually holds only at a discrete set of total
filling factors, including $\nu_T =1$.  To understand why these
concerns would be misplaced, it is useful to briefly review some
of the microscopic physics underlying the incompressible states whose occurrence
is responsible for the fractional quantum Hall effect.   For a pair of
interacting electrons confined to the lowest Landau level,
only one relative motion state is available for each
relative angular momentum.  This property means that the interaction
is characterized by a discrete set of energy scales, $V_{l}$, first identified
by Haldane\cite{haldanepspots} and known as Haldane pseudopotentials.
$V_{l}$ is the interaction energy of a pair of electrons with relative
angular momentum $l$; in the double-layer case, the pseudopotential
values for interlayer and intralayer interactions will differ.
Many of the largest charge gaps which occur
in the fractional quantum Hall regime can be understood in terms of
Haldane pseudopotentials.  For example, for double-layer systems
with nearby layers the largest energy scales should be the interlayer
and intralayer $l=0$ pseudopotentials.  It is easy to show that it
is possible to form many body-states in which pairs of particles
never have $l=0$ for $\nu_T \le 1$. This is not possible for $\nu_T > 1$.
Thus if $\nu_T = 1$, the energy to add a charge to the system is greater
than the energy to remove one, hence the charge gap.  For $\nu_T=1$
it is easy\cite{leshouches} to show that the only states which completely
avoid pairs with relative angular momentum $l =0 $ are,
up to  rotation in pseudospin space,
identical to the wavefunctions in Eq.~(\ref{eq:variational}).
This is the real reason why the variational wavefunctions
used above are accurate.    
In the following we describe low-energy states of the system as
spin-texture states with a position-dependent local pseudospin
orientation.  The property that the low-energy
states are completely specified by spin-textures, which we will use
frequently in following sections, depends on both the broken
symmetry of the ground state {\it and} on the existence of the charge gap.

The $U(1)$ symmetry leads to Eq.~(\ref{eq:xymod})
which defines an effective XY model
which will contain vortex excitations which interact logarithmically.
In a thin film of superfluid $^4$He, vortices interact logarithmically
because of the energy cost of supercurrents circulating around the
the vortex centers.
(In superconducting thin films the same logarithmic interaction appears
but is cut off on length scales exceeding the penetration depth.)
Here the same logarithmic interaction appears.  Microscopically
this interaction is due to
the potential energy cost (loss of exchange) associated with the
phase gradients (circulating pseudospin currents).
Hartree-Fock estimates\cite{Ilong}
indicate that $\rho_s$ and hence the Kosterlitz-Thouless
critical temperature are on the scale of 1K in typical samples.  Vortices in
the $\varphi$ field (`merons'\cite{Ilong})
are reminiscent of Laughlin's fractionally charged
quasiparticles but in this case carry charges $\pm\frac{1}{2} e$ and can
independently be
left- or right-handed for a total of four `flavors'.\cite{usPRL,Ilong}
Bound meron pairs with opposite vorticity are the lowest\cite{hfmeron} energy
charged excitations of the system.  The finite pseudospin stiffness
not only permits the presence of spontaneous pseudospin magnetization
but leads to a finite charge excitation gap
(even though the tunnel splitting is zero).  Thus the QHE
survives\cite{Ilong,hfmeron} the limit $\Delta_{\rm SAS} \longrightarrow 0$.

Since the `charge' conjugate to the phase $\varphi$ is the $z$
component of the pseudo spin $S^z$, the
pseudospin `supercurrent'
\begin{equation}
J =\frac{2\rho_s}{\hbar} {\bf \nabla}\varphi
\end{equation}
represents oppositely directed charge currents in each layer.  Below the
KT transition temperature, such current flow will be dissipationless
(in linear response) just as in an ordinary superfluid.  Likewise there
will be a linearly dispersing collective Goldstone mode as in a
superfluid.\cite{gapless,ahmz1,wenandzee,ezawa,usPRL,Ilong}

To reinforce the idea that this is not an ordinary superfluid or
superconductor, it is perhaps useful to rewrite the original variational
wave function as
\begin{equation}
|\psi\rangle = \prod_X\left[\frac{1}{\sqrt{2}}\left(1 + e^{i\varphi}
c_{X,\downarrow}^\dagger c_{X,\uparrow}\right)\right]|\psi_\uparrow\rangle,
\end{equation}
where
\begin{equation}
|\psi_\uparrow\rangle \equiv \prod_X c^\dagger_{X,\uparrow}|0\rangle
\end{equation}
is the fully up polarized spin state.  We now see that the analogy
to an excitonic insulator\cite{sfcond,macdp-hole} is closer than the
analogy to a superconductor.  Only
a gauge-neutral object (a particle bound to a hole) can condense and
propagate freely in a strong B field.  A phase gradient $\nabla\varphi$
causes a flow of these neutral objects which then constitutes a spin
current analogous to the charge current in a superconductor.
In analogy to the excitonic insulator case,
one can make a particle-hole transformation
in one of the layers to produce a formal resemblance to the BCS mean
field state:
\begin{equation}
|\psi\rangle = \prod_X\left[\frac{1}{\sqrt{2}}\left(1 + e^{i\varphi}
c_{X,\downarrow}^\dagger c_{X,\uparrow}^\dagger\right)\right]
|\psi_\uparrow\rangle,
\end{equation}
but it is then clear that one has a pairing between a particle and hole;
not between two particles.
It is worth remarking that there is in general
an exact mapping\cite{macdp-hole} between
double-layer electron-hole systems and two-component electron systems
in the limit of strong magnetic fields.  The Hamiltonian for a
double-layer electron system is mapped, up to a constant term,
to that for an electron-hole system if a particle-hole transformation
is made in one of the layers.  The filling factor of interest here,
$\nu_T =1$, corresponds to an electron-hole system with equal electron
and hole densities and, as explained above,
the spontaneous phase coherence state corresponds
to an excitonic superfluid state.  In the electron-hole case, the fact
that a broken symmetry ground state can occur had been appreciated
some years ago.\cite{exsuper}  Some of the recent advances in understanding
the physics associated with the spontaneous-phase-coherence ground state
in the electron-electron case, have important implications for the
electron-hole case which we will not explore at length here.

A finite tunneling amplitude $t$ between the
layers breaks the $U(1)$ symmetry
\begin{equation}
H_{\rm eff} = \int d^2r
\left[\frac{1}{2}\rho_s \vert\nabla\varphi\vert^2 - \frac{t}{2\pi\ell^2}
 \cos{\varphi}\right]
\label{eq:H_eff}
\end{equation}
by giving a preference to symmetric tunneling states.  This can be
seen from the tunneling Hamiltonian
\begin{equation}
H_{\rm T} = - t \int d^2r
\left\{\psi_\uparrow^\dagger ({\bf r}) \psi_\downarrow ({\bf r})
+ \psi_\downarrow^\dagger ({\bf r}) \psi_\uparrow ({\bf r})\right\}
\label{eq:tunnel}
\end{equation}
which can be written in the spin representation as
\begin{equation}
H_{\rm T} = - 2t \int d^2r S_x({\bf r}).
\end{equation}
(Recall that the eigenstates of $S_x$ are symmetric and
antisymmetric combinations of up and down.)

We can shed further light on the spontaneous symmetry breaking by
considering the tunneling Hamiltonian $H_{\rm T}$ in Eq.~(\ref{eq:tunnel})
as a weak perturbation. Naively, since particle number is separately
conserved in each layer for $t=0$, one might expect
\begin{equation}
\lim_{t\longrightarrow 0} \frac{1}{t} \left\langle\psi\left|H_{\rm
T}\right|\psi\right\rangle = 0.
\end{equation}
That is, one might expect that the first-order term in the perturbation
series for the energy due to $t$ to vanish.  Instead however we find that
the energy shifts {\it linearly\/} in $t$
\begin{eqnarray}
\lim_{t\longrightarrow 0} \lim_{A\longrightarrow\infty}
\frac{1}{tA} \left\langle\psi\left|H_{\rm T}\right|\psi\right\rangle
&=& \lim_{t\longrightarrow 0} \lim_{A\longrightarrow\infty}
\frac{1}{A}
\left\langle\psi\left|-\int d^2 r\, 2S_x({\bf r})\right|\psi\right\rangle\cr
&=& -m^x,
\end{eqnarray}
where $A$ is the system area, and $m^x$ is, by definition,
the magnetization which is
the system's order parameter.\cite{order-of-limits} If the interlayer
spacing $d$ is taken to be zero, one can readily show\cite{Ilong} that the
variational wave function in Eq.~(\ref{eq:variational}) is exact,
hence $\lim_{t\longrightarrow 0} m^x = 1$, and $t = \Delta_{SAS}/2$.
For finite $d$,
Eq.~(\ref{eq:variational}) is no longer exact and quantum
fluctuations will\cite{Ilong} reduce the magnitude of $m^x$ and
we must renormalize the hopping parameter $t$
appropriately.

As the layer separation $d$ increases, a critical point $d^*$ will be
reached at which the magnetization vanishes and the ordered phase is
destroyed by quantum fluctuations.\cite{usPRL,Ilong}  This is illustrated
in Fig.~(\ref{fig:qhe/no-qhe}).
For {\it finite\/} tunneling $t$, we will see below
that the collective mode becomes massive and quantum fluctuations will be
less severe.  Hence the phase boundary in Fig.~(\ref{fig:qhe/no-qhe})
curves upward
with increasing $\Delta_{\rm SAS}$.
For $\Delta_{SAS}=0$
the destruction of long-range order and the charge excitation gap
are intimately related and occur simultaneously at $d^*$ and zero
temperature.  For finite $\Delta_{SAS}$ the system
always has non-zero $m^x$ even in the phase with zero charge gap.

The effective Hamiltonian in Eq.~(\ref{eq:H_eff}) looks like the
sine-Gordon model which is known to have a finite temperature
Kosterlitz-Thouless phase transition.\cite{fradkin}  One might be
tempted to speculate then that the rapid collapse of the Arrhenius plots
of the dissipation at unexpectedly low temperatures is associated with
a true phase transition.  We believe however that these are merely rapid
cross-overs rather than true phase transitions because the phase $\varphi$
is compact.  The quantum states defined by $\varphi$ and $\varphi + 2\pi$
are microscopically identical.  The interpretation of $\varphi$ in a
sine-Gordon theory as an electrostatic potential for Coulomb charges
or as the surface height in the solid-on-solid model requires that
$\varphi$ and $\varphi + 2\pi$ be distinguishable states.  Hence what
we really have is an XY model in a symmetry breaking field which
has no true phase transition since the vortices are linearly 
confined.
[Note that the solid-on-solid model has no analog of open strings 
(see Sec. \ref{sec:chargedEX}) terminated
by vortices such as we have here.\cite{bookchap}]

\section{Charged Excitations}
\label{sec:chargedEX}
At filling factor $\nu=1$, there is an intimate connection between local
distortions of the pseudospin orientation and the local charge
density.\cite{Ilong}  A simple  example of this is provided by the fully
spin polarized $\nu=1$ state of Eq.~(\ref{eq:variational}).
Since the Landau level is filled, the charge density is uniform and the
Pauli principle forbids any (intra-Landau-level)
excitations which do not flip spins.  One can form a spin-flip
particle-hole pair at an energy cost of\cite{sondhi}
\begin{equation}
U_0 = \Delta_{\rm SAS} + \frac{e^2}{\epsilon\ell}
\left(\frac{\pi}{2}\right)^{\frac{1}{2}}
\end{equation}
where the first term is the tunneling energy necessary to flip the
pseudospin and the second represents the loss of Coulomb exchange energy
(here computed for the special case $d=0$).  The Coulomb exchange energy
gives a finite cost even in the absence of tunneling and this explains the
fact that the typical observed gap can be much larger than the tunnel
splitting.\cite{expamd,murphyPRL,mansour}   The simple spin flip is not
however the optimal charged-pair excitation.  From analogy with results of
Sondhi et al.\cite{sondhi} For the case of `real' spins, we know that
smooth local distortions of the pseudospins produce a charge 
density given by the remarkable formula\cite{sondhi,Ilong}
\begin{equation}
\delta \rho({\bf r}) = -\frac{\nu}{8\pi}\> \epsilon_{\mu\nu}\>
{\bf m}({\bf r})\cdot
\left[\partial_\mu {\bf m}({\bf r})\times
\partial_\nu {\bf m}({\bf r})\right]
\label{eq:3.240}
\end{equation}
where $\bf m$ is a unit vector giving the local pseudospin orientation.
$\delta \rho({\bf r})$ is exactly $\nu$ times the
Pontryagin index, or topological charge
density.\cite{sondhi,fradkin}
The density in Eq.~(\ref{eq:3.240}) can be viewed as the time-like component
of a conserved (divergenceless) topological `three-current'
\begin{equation}
j^\alpha = -\frac{\nu}{8\pi}\> \epsilon^{\alpha\beta\gamma}\>
\epsilon_{abc}\>
{m^a}({\bf r})
\partial_\beta {m^b}({\bf r})
\partial_\gamma {m^c}({\bf r}).
\label{eq:3.240a}
\end{equation}
Using the fact that $\bf m$ is a unit vector, it is straightforward to
verify that $\partial_\mu j^\mu = 0$.

`Skyrmion' (hedgehog) configurations of the order parameter carry net
charge $\pm 1$ and, when $\Delta_{\rm SAS} = 0$ and
$d=0$,\cite{sondhi,gapless} a skyrmion pair has an excitation energy which
is one-half of that of a simple spin-flip pair for the case of Coulomb
interactions between the electrons.  At finite $d$, the SU(2) symmetry is
lowered to U(1) and the cheapest charged excitations are composed of 
`merons' which are
essentially vortex solutions in which the local pseudospin winds
by $\pm 2\pi$ at infinity and tilts either up or down out of the XY
plane in the core region as shown in Fig.~(9) of I.
Integration of the charge density using Eq.~(\ref{eq:3.240}) shows that
vortices carry charge $\pm \frac{1}{2}$.  They are somewhat analogous
to Laughlin quasiparticles, however they differ considerably in that,
below the Kosterlitz-Thouless temperature, they are confined together
in vorticity neutral pairs by their logarithmic interaction.  The
cheapest object with a net charge is then a vortex-antivortex pair,
with each vortex
carrying charge $+\frac{1}{2}$ (or $-\frac{1}{2}$) for a total charge
of $+1$ (or $-1$).  The charge excitation cost can be estimated by
minimizing
\begin{equation}
E_{\rm pair} = 2E_{\rm mc} + \frac{e^2}{4\epsilon R} +
2\pi\rho_s\ln{\left[\frac{R}{R_{\rm mc}}\right]},
\end{equation}
where $E_{\rm mc}$ is the meron core energy\cite{hfmeron}
, and $R_{\rm mc}$ is the meron core size.
The optimal separation is given by\cite{Ilong}
$R_0 = e^2/(8\pi\epsilon\rho_s)$.
It is important for the discussion below that in typical double-layer
systems $\rho_s$ is much smaller than the microscopic energy scale
$e^2/\epsilon\ell$.
For $d/\ell =1$, $\rho_s = 6.19 \times 10^{-3} (e^2/\epsilon\ell)$
in the Hartree-Fock approximation and it is further renormalized downward
by quantum fluctuations.\cite{Ilong}  Typical values of $\rho_s$ for
double-layer systems are smaller than $5 \times 10^{-3} (e^2/\epsilon\ell)$ so
that $R_0/\ell$ will typically be larger than $\sim 8 \ell$.
The small values of the pseudospin stiffness allow the charged
pseudospin textures to be large, as required for the
validity of the long-wavelength description being employed here.

The introduction of finite tunneling amplitude destroys the U(1) symmetry
and makes the simple vortex-pair configuration extremely expensive.
To lower the energy the system distorts the spin deviations into a domain
wall or `string' connecting the vortex cores as shown in
Fig.~(\ref{fig:meron_string}).  The spins are oriented in the $\hat x$
direction everywhere except in the domain line region where they
tumble rapidly through $2\pi$.
The domain line has a fixed energy per unit length and so the vortices
are now confined by a linear potential corresponding to a fixed
`string tension' rather than being confined only 
logarithmically.  We can estimate the string tension by examining the
energy of a domain line of infinite length.  The optimal form
for a domain line lying along the $y$ axis is given by
\begin{equation}
\varphi({\bf r}) = 2 \arcsin{[\tanh{(x/\xi)}]},
\end{equation}
where the characteristic width of the string is
\begin{equation}
\xi = \left[\frac{2\pi\ell^2\rho_s}{t}\right]^\frac{1}{2}.
\end{equation}
The resulting string tension is\cite{gruner}
\begin{equation}
T_0 = 8 \left[\frac{t\rho_s}{2\pi\ell^2}\right]^\frac{1}{2} =
\frac{8 \rho_s}{\xi}.
\label{eq:tension_0}
\end{equation}
Provided the string is long enough ($R \gg \xi$), the total energy of
a segment of length $R$ will be well-approximated by the expression
\begin{equation}
E_{\rm pair}' = 2E_{\rm mc}' + \frac{e^2}{4\epsilon R} + T_0R.
\label{string_pair}
\end{equation}
The prime on $E_{mc}$ in Eq.~(\ref{string_pair}) indicates
that the meron core energy can
depend on $\Delta_{SAS}$.  $E_{\rm pair}'$ is minimized
at $R=R_{0}' \equiv \sqrt{e^2/4\epsilon T_0}$ where it has the value
\begin{equation}
E^*_{pair} = 2E_{\rm mc}' + \sqrt{e^2 T_0/\epsilon}.
\label{string_pair_eng}
\end{equation}
Note that apart from the core energies, the charge gap at fixed
layer separation (and hence fixed $\rho_s$) is
$\propto T_0^{1/2} \propto t^{1/4} \sim \Delta_{SAS}^{1/4}$,
which contrasts with the case of free electrons, for which
the charge gap is linearly proportional to $\Delta_{SAS}$.  Note that because
the exponent $1/4$ is so small, there is an extremely
rapid initial increase in the charge gap as tunneling is turned on.

The crossover between the meron-pair pseudospin texture
which holds for $t \equiv 0$ and the domain line string pseudospin
texture described above occurs at a finite value of $t$ which we
can estimate by the following argument.  For $R_{0}' > R_0$ the
vortices are already bound by the logarithmic attraction due to
the gradient energy before the linear attraction due to the hopping
becomes important at larger separations.  In this regime
tunneling does not play an important role in determining the nature
of the lowest energy charged pseudospin texture.  As $t$ increases
$R_{0}' \propto t^{-1/4}$ decreases and will eventually reach $R_0$
which is, of course,  independent of $t$.  Since
\begin{equation}
\frac{R_{0}'}{R_0} =
\left(\frac{ 2 \pi^2 \rho_s}{e^2/\epsilon \xi} \right)^{1/2}
= \frac{\pi \xi}{4 R_{0}'},
\label{eq:crossover}
\end{equation}
the characteristic width of the domain line becomes comparable to
$R_{0}'$ in the same range of $t$ values where $R_{0}'$ and $R_0$
become comparable.  We may conclude that the nature of the
charged pseudospin texture crosses over directly from the
meron pair form to the finite length domain line string form for
$\rho_s/(e^2/\epsilon\xi) \sim 1/25$,
or equivalently for $t \sim t_{\rm cr}$ where
\begin{equation}
t_{\rm cr} = 4 \times 10^3 {\biggl[\frac{\rho_s}{e^2/\epsilon\ell}\biggr]}^{3}
\frac{e^2}{\epsilon\ell}.
\label{eq:tcross}
\end{equation}
The crossover tunneling amplitude is thus typically smaller than
$ 5 \times 10^{-4} (e^2/\epsilon\ell)$.   Typical tunneling amplitudes in
double-layer systems are smaller than $ \sim 10^{-1} (e^2/\epsilon\ell)$
and can be made quite small by adjusting the barrier material 
or making the barrier wider.  Nevertheless, it seems likely that
$t$ will be larger than $t_{\rm cr}$ except for samples which are carefully
prepared to make $t$ as small as possible.  As $t$ increases beyond
$t_{\rm cr}$, $R_{0}'$ will continue to decrease.  When $R_{0}'$ becomes
comparable to the microscopic length $\ell$, the description given
here will become invalid and the lowest energy charged excitations
will have single-particle character.  However, the domain-wall string
picture of the charged pseudospin texture has a very large
range of validity since $R_{0}' \propto t^{-1/4}$ decreases
very slowly with increasing $t$.  Writing $R_{0}'
\sim (e^2 / 8\epsilon \pi \rho_s)
(t_{\rm cr}/t)^{1/4}$ we find that $R_{0}' \sim \ell$ only for
$ t \sim 10^{-2} [(e^2/\epsilon\ell)^2/\rho_s ]$.  Using typical values of
$\rho_s$ we see that the charged excitation crosses over to single
particle character only when the hopping energy $t$ becomes
comparable to the microscopic interaction energy scale.  The various
regimes for the charge excitations of double-layer systems are
summarized in Table~\ref{table:regimes}.  Almost
all typical double-layer systems lie within the regime of the
domain-wall-string pseudospin texture charge excitation.

We should emphasize that all the discussion of charged excitations
above assumes that the meron core energies do not make a
dominant contribution to the charge excitation energies and
that meron core sizes are small compared to the overall size
of the quasiparticles.  Recent calculations\cite{breyetalhf} have demonstrated
that these conditions are never well satisfied when
the Hartree-Fock approximation is used to approximate the
charged excitations.  The physical pictures summarized
still have some qualitative validity, however.  The Hartree-Fock
approximation neglects quantum fluctuation effects which reduce
both the spin-stiffness and the order parameter, tending in both
cases to increase the size of the quasiparticles and increase
the appropriateness of the pictures presented here.

\section{Parallel Magnetic Field}
\label{sec:parallelB}

Murphy et al.\cite{murphyPRL,jpebook} and Santos et al.\cite{mansour}
have shown that the charge gap in double layer
systems is remarkably sensitive to the application of relatively weak magnetic
fields $B_\parallel$, oriented in the plane of the 2D electron gas.
Experimentally this field component is generated by
slightly tilting the sample relative to the magnetic field orientation.
Tilting the field (or sample) has
traditionally been an effective method for identifying effects due to
(real) spins because orbital motion in a single-layer 2DEG system
is primarily\cite{nickila} sensitive to $B_\perp$, while
the (real) spin Zeeman splitting is proportional to the full magnitude of
$B$.  Adding a
parallel field component will tend to favor more strongly
spin-polarized states.
For the case of the double layer $\nu =1$ systems studied by
Murphy et al., \cite{murphyPRL} the
ground state is known to already be an isotropic ferromagnetic state  of
the {\it true spins} and the addition of a parallel field would not,
at first glance,
be expected to influence the low energy states since they are already
fully spin-polarized. (At a fixed Landau level filling factor
$B_\perp$ is fixed and so both the total $B$
and the corresponding Zeeman energy increase with tilt).
Nevertheless experiments\cite{murphyPRL} have shown that
these systems are very sensitive to $B_\parallel$.  The activation energy
drops rapidly (by factors varying from two up to an order-of-magnitude
in different samples) with increasing $B_\parallel$.
At $B_\parallel = B_\parallel^\ast$ there appears to be
a phase transition to a new state whose activation gap is approximately
independent of further increases in $B_\parallel$.

The effect of $B_\parallel$ on the {\it pseudospin\/} system
can be visualized in
two different pictures.  In the first picture
we use a gauge in which ${\bf B}_\parallel =
{\bf\nabla}\times {\bf A}_\parallel$ where
${\bf A}_\parallel = B_\parallel (0,0,x)$.
In this gauge the vector potential points in the $\hat z$ direction
(perpendicular to the layers) and varies with position $x$ as one
moves parallel to the layers.
As an electron tunnels from one layer to the other it moves along the
direction in which the vector potential points and
so the tunneling matrix element acquires a position-dependent phase
$t\rightarrow t~e^{iQx}$ where $Q=2\pi /L_\parallel$ and $L_\parallel =
\Phi_0/B_\parallel d$ is the length associated with one flux quantum
$\Phi_0$ between the layers [defined in Fig.~(\ref{fig:fig2})].
This modifies the tunneling Hamiltonian to $H_T=-\int
d^2r~{\bf h}({\bf r})\cdot {\bf S}({\bf r})$ where ${\bf h}({\bf r})$
`tumbles': {\it i.e.},
${\bf h}({\bf r})=2t~(\cos{Qx},\sin{Qx},0)$.  The effective XY
model now becomes
\begin{equation}
H=\int d^2r~\Biggl\{ \frac{1}{2}~\rho_s\vert {\bf\nabla}\varphi\vert^2 -
\frac{t}{2\pi\ell^2}~\cos{[\varphi ({\bf r}) - Qx]}\Biggr\} ,
\label{eq:140smg}
\end{equation}
which is precisely the Pokrovsky-Talapov (P-T) model\cite{bak} and has a
very
rich phase diagram.  For small $Q$ and/or small $\rho_s$ the phase obeys
(at low temperatures)
$\varphi
({\bf r})\equiv Qx$; the moment rotates commensurately with the pseudospin
Zeeman
field.  However, as $B_\parallel$ is increased,
the local field tumbles too rapidly and a continuous
phase transition to an incommensurate state with broken
translation symmetry occurs.  This is because
at large $B_\parallel$ it costs too much exchange
energy to remain commensurate and the system rapidly gives up the tunneling
energy in order to return to a uniform state ${\bf\nabla}\varphi\approx 0$
which becomes independent of $B_\parallel$.  As explained in further detail
below we\cite{usPRL}
find that the phase transition occurs at zero temperature for
\begin{equation}
B_\parallel^\ast = B_\perp ~ (2 \ell / \pi d) (2 t / \pi \rho_s)^{1/2}.
\label{eq:critparfield}
\end{equation}
Using the parameters of the samples
of Murphy et al. \cite{murphyPRL} and neglecting quantum fluctuation
renormalizations of both $t$ and $\rho_s$ we find that
the critical field for the transition is
$ \approx 1.6 {\rm T}$ which is within a factor of two
of the observed value\cite{murphyPRL}.
Note that the observed value $B_\parallel^\ast =0.8{\rm T}$
corresponds in
these samples to a large value for $L_\parallel$: $L_\parallel /\ell\sim
20$ indicating that the transition is highly collective in nature.
We emphasize again that these very large length scales are possible in
a magnetic field  only
because of the interlayer phase coherence in the system associated with
condensation of a {\it neutral\/} object.

Having argued for the existence of the commensurate-incommensurate
transition, we must now connect it to the experimentally observed
transport properties.
In the commensurate phase, the order parameter tumbles more and more rapidly
as $B_\parallel$ increases.  As we shall see below,
it is this tumbling which causes
the charge gap to drop rapidly.  In the incommensurate phase the state
of the system is approximately independent of $B_\parallel$ and this
causes the charge excitation gap to saturate at a fixed value.

Recall that in the presence of tunneling, the cheapest charged excitation
was found to be a pair of vortices of opposite vorticity and like charge
(each having charge $\pm 1/2$) connected by a domain line
with a constant string tension.  In the absence of $B_\parallel$ the
energy is independent of the orientation of the string.  The effect of
$B_\parallel$ is most easily studied by changing variables to
\begin{equation}
\theta({\bf r}) \equiv \varphi({\bf r}) - Qx.
\end{equation}
This variable is a constant in the commensurate phase but not in
the incommensurate phase.  In terms of this new variable,
the P-T model energy is
\begin{equation}
H=\int d^2r~\Biggl\{ \frac{1}{2}~\rho_s [(\partial_x\theta + Q)^2
+ (\partial_y\theta)^2]
- \frac{t}{2\pi\ell^2}~\cos{\theta}\Biggr\} .
\label{eq:140asmg}
\end{equation}
We see that $B_\parallel$ defines a preferred direction in the problem.
Domain walls will want to line up in the $y$ direction and contain
a phase slip of a preferred sign ($-2\pi$ for $Q>0$) in terms of
the field $\theta$.  Since the extra term induced by $Q$ represents a total
derivative, the optimal form of the soliton solution is unchanged.  However the
energy per unit length of the soliton,
which is the domain line string tension,
decreases linearly with $Q$ and hence $B_\parallel$:
\begin{equation}
T = T_0\left[1 - \frac{B_\parallel}{B_\parallel^*}\right],
\end{equation}
where $T_0$ is the tension in the absence of parallel B field
given by Eq.~(\ref{eq:tension_0})
and
$B_\parallel^*$ is the critical parallel field at which the string tension
goes to zero.\cite{commentrenorm}  We thus see that by tuning $B_\parallel$
one can conveniently control the `chemical potential' of the domain lines.
The domain lines condense and
the phase transition occurs when the string tension becomes negative.

Recall that the charge excitation gap is given by the energy of a vortex
pair separated by the optimal distance $R_0 = \sqrt{e^2/4\epsilon T}$.
\ From Eq.~(\ref{string_pair}) we have that the energy gap for the
commensurate state of the phase transition is given by
\begin{eqnarray}
\Delta &=& 2 E_{\rm mc}' + \left[e^2 T/\epsilon\right]^\frac{1}{2}\cr
&=& \Delta_0 + \sqrt{e^2T_0/\epsilon}
\left[1-\left(\frac{B_\parallel}{B_\parallel^*}
\right)
\right]^\frac{1}{2}.
\end{eqnarray}
As $B_\parallel$ increases, the reduced string tension allows the Coulomb
repulsion of the two vortices to stretch the string and lower the energy.
Far on the incommensurate side of the phase transition
the possibility of interlayer tunneling becomes irrelevant.  From
 the discussion of the previous section it follows that the ratio
of the charge gap at $B_{\parallel} = 0 $ to the charge
gap at $B_{\parallel} \to \infty$ should be given approximately by
\begin{equation}
\frac{\Delta_0}{\Delta_{\infty}} = (t/t_{\rm cr})^{1/4} \approx
{(e^2/\epsilon\ell)^{1/2} t^{1/4}}{ 8 {\rho_s}^{3/4}}.
\label{eq:gapratio}
\end{equation}
Putting in typical values of $t$ and $\rho_s$ gives gap ratios $\sim
1.5-7$  in agreement with experiment.  According to the discussion of the
previous section, gap ratios as large as $\sim (t_{max}/t_{\rm cr})^{1/4}
\sim 0.07 (e^2/\epsilon\ell) / \rho_s$, can be expected in the regime
where the pseudospin texture picture applies. Here $t_{max}$ is the
hopping parameter at which the crossover to  single-particle excitations
occurs.  Thus gap ratios as large as an order of magnitude are easily
possible.  Of course, all the discussion here  neglects orbital effects
(electric subband mixing)
 within each of the electron gas layers, and these will always become
important at sufficiently strong parallel fields.

It should be emphasized that only this highly collective picture involving
large length scale distortions of topological defects can possibly explain
the extreme sensitivity of the charge gap to small tilts of the B field.
Recall that at $B_\parallel^*$ the tumbling length $L_\parallel$ is much
larger than the particle spacing and the magnetic length.  Simple
estimates of the cost to make a local one-body type excitation (a
pseudospin-flip pair for example) shows that the energy decrease due to
$B_\parallel$ is extremely small since
$\ell/L_\parallel$ is so small.  As we will see in Sec.
\ref{sec:exact_diag} numerical exact-diagonalization calculations on small
systems confirm the existence of this  phase transition and show that the
fermionic excitation gap drops to a much smaller value in the
incommensurate phase.

We now discuss the commensurate-incommensurate phase
transition from the microscopic point of view.
At $d=0$, the $B_\parallel = 0$ Landau-gauge many-body ground state
wavefunction is a single Slater determinant in which
the single-body states are the symmetric linear combination
of two single-layer states with the same guiding center.
This is the state represented by Eq.~(\ref{eq:variational}) with
$\varphi=0$.
Phase coherence is established (either spontaneously or in this
case) by tunneling between
single-layer states with the same guiding center.
For many purposes this state is still a good approximation to
the ground state at finite $d$ since it
optimizes the tunneling energy and has good correlation energy; an
electron in one layer automatically sees an
exchange-correlation hole in the other layer at the
same place.  (It would remain an exact ground state at finite $d$
in the absence of
interactions.) From a microscopic point of view it is the good interlayer
correlations of states with phase coherence which leads to the
broken symmetry in the absence of tunneling.

To see the effect of a parallel B field, it is convenient to choose
a new Landau gauge for the perpendicular field
${\bf A}_\perp = -B_\perp (y,0,0)$.
In this gauge, a parallel field giving rise to
$t \to t \exp (i Q x) $ causes tunneling to couple
states in the two layers whose wavevectors differ by $Q$ and whose
guiding centers therefore differ by $Q\ell^2$.\cite{tunneldiff}
Thus, for non-interacting electrons the exact ground state in a parallel field
is one in which the exchange-correlation hole is not directly opposite its
electron but rather is shifted away by $\ell^2 Q \hat y$
as the {\bf B} field tilts in the $\hat x$ direction (i.e., the displacement
is perpendicular to the direction of the in-plane field):
\begin{equation}
|\psi_Q\rangle = \prod_Y\left\{c^\dagger_{Y,\uparrow} +
c^\dagger_{Y+Q\ell^2,\downarrow}\right\}|0\rangle.
\end{equation}
This state maintains all of its
tunneling energy but rapidly loses interlayer
correlation energy as the field tilts.
At large tilt it is better to give up on the tunneling by shifting the two
layers relative to each other to put the correlation hole back next to its
electron.

This shift can be seen to be the change from commensurate to
incommensurate states discussed above.  A straightforward  
computation shows that the commensurate state has
the pseudospin tumbling
\begin{equation}
\langle\psi_Q|\psi^\dagger_\uparrow({\bf r})
\psi_\downarrow({\bf r})| \psi_Q\rangle = e^{-\frac{1}{4}Q^2\ell^2}
e^{-iQx},    
\end{equation}
while the pseudospin is constant in the incommensurate phase.

All of our discussion of the phase transition in a parallel field
has been based on mean-field theory.  Close to the phase transition,
thermal fluctuations will be important.  At finite temperatures
there is no strict phase transition at $B_\parallel^\ast$ in the
the P-T model.  However there is a finite temperature KT phase transition
at a nearby $B_\parallel > B_\parallel^\ast$.  At finite temperatures
translation symmetry is restored\cite{bak} in the incommensurate phase
by means of dislocations in the domain string structure.
Thus there are two separate KT transitions in this system,
one for $t=0$, the other for $t\neq 0$ and $B_\parallel > B_\parallel^\ast$.
Recently Read\cite{read2layer} has studied this model at finite
temperatures in some detail and has shown
that at the critical value of $B_\parallel$ there should be
a square-root singularity in the charge gap on both sides
of the transition.  The existing data does
not have the resolution to show this however.  At zero-temperature
the commensurate-incommensurate\cite{mpafcomment,read2layer}
phase transition must be treated quantum mechanically.
It is necessary to account for the
world sheets traced out by the time evolution of the strings which
fluctuate into existence due to quantum zero-point motion.  Read also points
out that the inevitable random variations in the tunneling
amplitude with position,
which we have not considered at all here, cause a relevant perturbation.

\section{Collective Modes and Response Functions}
\label{sec:collective_modes}

In this section we will discuss the charge neutral collective
excitations of double-layer systems and some physically
important response functions which have poles at the collective
excitation energies.  In the pseudospin language, the collective excitations
are spin-waves in which the pseudospin precesses around its ground state
orientation. We will thus need to
enlarge our description of the system by allowing the
pseudospin texture to have orientations out of the $\hat x-\hat y$ plane.
This requires that we generalize from the $XY$-model description
of the system employed in previous sections, to the more complete
anisotropic non-linear sigma model description which we have discussed at
length in Ref.[~\onlinecite{Ilong}] (I).  (Actually this generalization is also
required if want to render the physics of the meron cores.)  States
of the system are characterized by a pseudospin texture
function, $\hat m (\bf r)$ which specifies
the space dependence of the pseudospin orientation.  The energy of a
pseudospin texture is given by the following functional
(where we retain only the leading terms in number of derivatives):
\begin{equation}
E[\hat{m}({\bf r})] = \int d^{2}{\bf r} \left[\beta
(m_{z}^{2}({\bf r})) +\frac{\rho_{s}}{2}
|\nabla m_{\perp}|^{2}-nt\; m_{x}({\bf r})\right]
\label{eq:glfunctional}
\end{equation}
where ${\bf \hat m} (\bf r) \cdot \hat m (\bf r)  \equiv 1$ and
$ m_{\perp} $ is the projection of the unit vector onto the
$\hat x- \hat y$ plane.   The energy is with respect to the
ground state in the absence of tunneling.
Here $n = (2 \pi \ell^2)^{-1}$ is the total electron density.
This energy functional is expected to be accurate for states
with spin-textures which vary slowly on a microscopic scale.
If the pseudospin is confined to the $\hat x-\hat y$ plane, its
orientation is specified by its azimuthal angle and the energy
functional reduces to the $XY$ functional used above.  This
Ginzburg-Landau energy functional differs from the one discussed
in Ref.[~\onlinecite{Ilong}] only through the addition of the tunneling
term.  The form of the functional follows from symmetry considerations
and the parameters can be considered as phenomenological constants.
In Ref.[\onlinecite{Ilong}], Eq.~(\ref{eq:glfunctional}) was derived
microscopically using a Hartree-Fock approximation, and
explicit expressions for the coefficients were obtained,
which become exact in the limit, $d \to 0$.  We now generalize the
discussion of collective modes given there to the case where tunneling
occurs between the layers.

\subsection{Collective Modes with Tunneling: $B_{\parallel} = 0$}

In the presence of inter-well tunneling, the pseudospin-orientation
in the ground state is constant and points in the $\hat x$ direction.
To calculate response functions we use the equations of motion
derived in Ref.[~\onlinecite{Ilong}].  We will assume that for low-lying
excitations the pseudospin orientation is always close to the $\hat x$
direction.
Then the equation of motions of the pseudospin texture in Fourier space is:
\begin{eqnarray}
\frac{dm_{y}({\bf q})}{dt}&=& -\frac{4\pi \ell^2 }{\hbar }
\frac{dE[m]}{dm_{z}(-{\bf q})}\\
\frac{dm_{z}({\bf q})}{dt}
&=& \frac{4\pi \ell^2 }{\hbar } \frac{dE[m]}{dm_{y}(-{\bf q})}
\end{eqnarray}
In the rest of this section we will use $\ell$ as the unit of length.
It is possible to linearize these equations if the pseudospin orientation
is close to the $\hat x$ direction by letting
$m_x = 1 - (m_y^2+m_z^2)/2 + \cdots$,
and dropping terms higher than first order in $m_y$ and $m_z$.  For
the double-layer system $m_y$ is proportional to the current flowing locally
between layers and $m_z$ is proportional to the difference between the
local densities in the two layers.  To calculate the response functions
of interest we add terms to the energy functional corresponding to pseudospin
Zeeman fields in the $\hat y$ and $\hat z$ directions, $h_y$ and $h_z$.
Physically $h_y$ corresponds to a perturbation in which an
imaginary term is added to the
tunneling amplitude and $h_z$ corresponds to a bias
potential between the two layers.
Linearizing and adding the Zeeman terms we find that
\begin{eqnarray}
\frac{dm_{y}({\bf q})}{dt} &=& -\frac{4\pi}{\hbar} (2\beta
+tn) m_{z}({\bf q}) + h_z\\
\frac{dm_{z}({\bf q})}{dt}
&=& \frac{4\pi}{\hbar } (tn+\rho_{s}q^{2})m_y({\bf q}) - h_y .
\end{eqnarray}

There is a similarity between these equations of motion
and those of the
Josephson effect, which is connected to the similarities between
interlayer phase coherence and superconductivity mentioned
previously.\cite{josephsonnote1}
The current across a Josephson junction between two superconductors
is proportional to $\sin(\phi)$ where $\phi$ is the
difference in the phase of the order parameter across the junction.
The current between layers in a double-layer system is similarly
proportional to $m_y$, {\it i.e.} proportional to $\sin (\phi)$ where
$\phi$ in the double-layer case specifies orientation of the component
of the pseudospin in the $\hat x- \hat y$ plane.  [In the linearized
case we are discussing we can equate $\phi$ and $\sin (\phi)$.]
More physically $\phi$ in the double-layer case specifies the difference
between the phase-coherence angle and the phase angle
of the interlayer tunneling amplitude.  In the case of the Josephson effect
\begin{equation}
\frac{d \phi}{d t} = \frac{2 e V}{\hbar},
\label{eq:jeffect}
\end{equation}
where $V$ is the potential drop across the junction.  Thus the equations
of motion for the phase angles in the Josephson effect case and
in the present case differ because of the presence of the term
proportional to $(2 \beta + t n) m_z({\bf q})$; a Josephson effect could
be achieved if the tunneling current were somehow extracted from the
double-layer system sufficiently
quickly to prevent any difference in the charge densities of the
two-layers from building up, {\it i.e.,} if $m_z$ where identically
zero.  In the case of double-layer systems, unlike the case of a
Josephson junction between two superconductors, it seems to
be impossible to do this.
The results we derive below are for an isolated double-layer
system and the resonance frequencies we
obtain are analogous to the Josephson plasma oscillations in an
isolated Josephson junction.

Solving for the pseudospin magnetization induced by
Zeeman fields with frequency $\omega$
we obtain the following result for the pseudospin
response tensor:
\begin{equation}
\left(\begin{array}{c}
  m_{y}\\
  m_{z}\end{array}\right) =\frac{\hbar}{E_{cm}^{2}-(\hbar \omega)^{2}}
\left(\begin{array}{cc}
  -i\omega & 8\pi \beta +2t\\
  -(2t+4\pi q^{2}\rho_{s}) & -i\omega\end{array}\right)\;
\left(\begin{array}{c}
  - 4 \pi h_{z}\\
  4 \pi h_{y}\end{array}\right)
\label{eq:responsetensor}
\end{equation}
where the collective mode energy is given by
\begin{equation}
E_{cm} = {\big[ (2t + 8 \pi \beta) (2t + 4 \pi q^2 \rho_s) \big]}^{1/2}.
\label{eq:cmbpeq0}
\end{equation}
The linearly dispersing collective mode of the $t \to 0$ case acquires a
gap because of the lifting of the $U(1)$ symmetry.

All components of this response tensor have poles at the collective
mode energies.  The response of the charge density difference to
a time-dependent inter-layer bias potential is given by
\begin{equation}
\chi_{zz}\equiv\frac{m_{z}}{h_{z}}=\frac{(4 \pi \hbar ) 2 t+4\pi
q^{2}\rho_{s}}{E_{cm}^{2}-(\hbar\omega)^{2}}
\label{eq:chizz}
\end{equation}
Using the continuity equation we can evaluate the corresponding
conductivity, the response in oppositely directed electric
currents to oppositely directed electric fields in the two-layers:
\begin{equation}
\sigma_{zz}(q, \omega) = e^2 \omega \chi_{zz} (q,\omega) / i q^2
\label{wq:sigzz}
\end{equation}
For $t \to 0$ the real part of the conductivity has a $\delta$-function
peak at zero frequency leading to spin-channel superfluidity.\cite{Ilong}
In the presence of interlayer tunneling the $\delta$-function peak
is shifted to $\omega = 2 t /\hbar $ and the superfluid behavior is lost.

In the static ($\omega \to 0$) long-wavelength ($q \to 0$) limit
$\chi_{zz}$ approaches a constant:
\begin{equation}
\chi_{zz}(\omega =0;q=0)= \frac{1}{\left(8\pi \beta +2t\right)}.
\label{eq:chizzstatic}
\end{equation}
A constant static interlayer
bias potential will\cite{jungwirth} simply tilt
the pseudospin orientation slightly out of the $\hat x-\hat y$ plane.
The effect of interlayer tunneling is to favor smaller tilts.
On the other hand
\begin{equation}
\chi_{yy}(\omega =0,q=0) =\frac{1}{2t}.
\label{eq:chiyystatic}
\end{equation}
If the tunneling amplitude goes to zero there is no restoring force
for rotations of the pseudospin in the $\hat x - \hat y$ plane, the
addition of an infinitesimal imaginary tunneling amplitude
will shift the
pseudospin orientation from the $\hat x$ direction to the $\hat y$
direction, and $\chi_{yy}(\omega=0,q=0)$ diverges.

\subsection{Collective Modes with Tunneling: Commensurate State}

For $B_{\parallel} =0$ we have been able to calculate collective
mode energies and linear response functions by linearizing the
non-linear sigma  model energy functional around the ground state
pseudospin orientation.  For $B_{\parallel} \ne 0$, the
ground state pseudospin orientation in the commensurate state
rotates with the pseudospin Zeeman field and it is necessary to
linearize the non-linear sigma  model energy functional around the rotating
pseudospin texture.  In anticipation of our needs for the case
of the incommensurate state we allow an arbitrary rate of rotation
for the rotating frame pseudospin function:
\begin{eqnarray}
\tilde{m}_{x} &=& m_{x}\cos{Px}-m_{y}\sin{Px} \\
\tilde{m}_{y} &=& m_{x}\sin{Px}+m_{y}\cos{Px}.
\label{eq:rotate}
\end{eqnarray}
In order to be able to treat the problem analytically, we limit ourselves
to the case of a position-independent tumbling rate $P$. 
The Ginzburg-Landau energy functional, expressed in terms of the rotating
frame pseudospins, is:
\begin{eqnarray}
E &=& \int d^{2} r\; \Biggl[\beta m_{z}^{2}+
\frac{\rho_{s}}{2} |\nabla
\tilde{{\bf m}}_{\perp}|^{2}+\frac{\rho_{s} P^{2}}{2}
|\tilde{{\bf m}}_{\perp}|^{2}\nonumber\\
&& -tn [\tilde{m}_{x}
\cos (P-Q)x+\tilde{m}_{y} \sin (P-Q)x] \nonumber\\
  && + \rho_{s}P\hat{z}\cdot\left(\frac{\partial\tilde{{\bf m}}_{\perp}}
{\partial x}\times
\tilde{{\bf m}}_{\perp}\right)\Biggr].
\label{eq:rotatedgl}
\end{eqnarray}
If we allowed only
translationally invariant spin-textures in the rotating frame
for which the lowest energy occurs,  this energy functional would
be minimized for small parallel fields by
choosing $P=Q$ to obtain an energy per area of $\rho_s Q^2/2 - t n$.
This is the commensurate state.
At large parallel fields the energy functional would be minimized by
the state with $P=0$, which has the same energy as if no tunneling
were present.  These two states cross in energy when $ Q^2 \xi^2  = 2$.
However, as we detail in the following section, the ground state at
high parallel fields can lower its energy further by
breaking translational symmetry.  The commensurate state is the ground
state only for $Q < Q_c$ where $Q_c^2 \xi^2 = 16 / \pi^2$.  $Q =Q_c$
for $B_{\parallel}= B_{\parallel}^*$.

In order for the linearization of this functional to be valid we
must assume we can   
choose $P$ so that $\tilde{m}_{x}$ is close to $1$ {\it everywhere},
both for the ground state and for the collective excitations in which we
are interested.  Assuming this to be the case we can apply periodic
boundary conditions in the thermodynamic limit and drop the total
derivative term in the energy functional.
For the commensurate state the linearized energy functional simplifies to
the following Fourier space version:
\begin{equation}
E[\tilde{m}_y,m_z] = A\;(\frac{\rho_s Q^2}{2} - t n) + \sum_{{\bf q}}
\left(\beta
+\frac{tn}{2}-\frac{\rho_{s}Q^{2}}{2}\right)
|m_{z}({\bf q})|^{2}+\left(\frac{\rho_{s}}{2} q^{2}+ \frac{tn}{2}\right)
|\tilde{m}_{y}|^{2}.
\label{eq:glengcs}
\end{equation}
The first term on the right-hand-side of Eq.~(\ref{eq:glengcs}) is the
ground state energy of the commensurate state.
The fact that it becomes
positive at large $Q$ implies that the commensurate
state eventually becomes unstable.  As discussed in previous sections
we expect the state to become unstable to the introduction of phase-slips
or solitons, which involve only the planar portion of the pseudospin
texture, when $ \rho_s Q^2 > 16 tn/\pi^2$.  We note from
Eq.~(\ref{eq:glengcs}) that if $\beta$ is small enough the
ground state will become unstable with respect to textures with
$m_z \ne 0$ before the solitons are introduced.  The requirement for
the applicability of
the scenario we introduced previously based on the $XY$ language
is that $ \beta \ell^2 > t (16 - \pi^2)/(4 \pi^3)$.  From estimates of
$\beta$ in Ref.[\onlinecite{Ilong}], it is clear that for present
double-layer samples this condition is secure.  Nevertheless, the
possibility exists that the behavior in parallel fields could be
quite different from that described here, for double-layer systems
with a layer spacing much smaller than what is achievable at present.
We have not explored this regime in detail and will assume in
what follows that $2 \beta + tn -\rho_s Q^2$ is positive in the
incommensurate state.
The effect of $B_{\parallel}$ is then to reduce the collective mode
energy, just as the charged excitation energies are lowered.
Solving the equations of motion with the linearized energy functional
of the commensurate state we find that
\begin{equation}
E_{\rm cm}  = \left[(2t + 8\pi\beta - 4\pi\rho_{s}Q^{2})
(2t + 4\pi\rho_{s}q^{2})\right]^{1/2}
\end{equation}
This result agrees with collective mode energies calculated in the
Hartree-Fock approximation\cite{cotesoliton} if the Hartree-Fock approximation
is used for the pseudospin stiffness.

\subsection{Collective Modes with Tunneling: Incommensurate State}

For $B_{\parallel} > B_{\parallel}^*$ it becomes energetically
favorable to break translational symmetry and
introduce phase slips, or solitons,  into the pseudospin texture.
In the ground state the phase slips are periodic\cite{bak,cotesoliton}
with a period $L_s$
which is determined by minimizing the energy.  In general then we choose
to work with pseudospins in a (again uniformly) 
rotating reference frame with
\begin{equation}
P=\left\lbrace\begin{array}{cc}
  Q & Q<Q_{c}\\
  Q-\frac{2\pi}{L_{s}}=Q-Q_{s} & Q\geq Q_{c}\end{array}\right.
\end{equation}
where $Q = 2 \pi/ L_{\parallel} \propto B_{\parallel}$.
For $P \ne Q$ the linearized energy functional is obtained by
setting $\tilde m_{\perp}^2 = 1 - m_z^2$, $\tilde m_x
= 1 - (\tilde m_y^2 +m_z^2)/2$, and dropping terms higher than
second order in $\tilde m_y$ and $m_z$.  We obtain
\begin{eqnarray}
E &=& \frac{A \rho_s P^2}{2} + \int d^{2}r\; \Biggl[\left(\beta
-\frac{\rho_{s}P^{2}}{2}\right) m_{z}^{2}
+\frac{\rho_{s}}{2} |\nabla\tilde{m}_{y}|^{2}\nonumber\\
 &&+\frac{tn}{2}\, (\tilde m_{y}^{2}+m_{z}^{2}) \cos(P-Q)x\nonumber\\
 &&-tn\, \tilde{m}_{y} \sin(P-Q)x\nonumber\\
 &&-\rho_{s}P\left(\frac{\partial \tilde{m}_{y}}{\partial x}\right)\Biggr] .
\end{eqnarray}
To be consistent with the linearization assumptions, $P$ must be
chosen so that $\frac{\partial \tilde{m}_{y}}{\partial x}$ integrates
to $0$ and this term can be dropped.

In Fourier space
\begin{eqnarray}
E &=& \frac{A\rho_{s}P^{2}}{2} +
\sum_{\bf p} \Biggl\{\left(\beta
-\frac {\rho_{s}P^{2}}{2}\right)
|\tilde{m}_{z}({\bf p})|^{2}+\frac{\rho_{s}q^{2}}{2}
|\tilde{m}_{y}({\bf p})|^{2}\Biggr\}\nonumber\\
 &&+\frac{tn}{4} \sum_{\bf p}\Biggl\{\tilde{m}_{y}({\bf p})
\biggl[\tilde{m}_{y}(-{\bf p}+(P-Q)\hat{x})+\tilde{m}_{y}
(-{\bf p}+(Q-P)\hat{x})\biggr]\nonumber\\
 &&\phantom{+\frac{tn}{4} \sum_{{\bf p}}}+m_{z}({\bf p}) \biggl[
m_{z}(-{\bf p}+(P-Q)\hat{x})+
m_{z}(-{\bf p}+(Q-P)\hat{x})\biggr]\Biggr\}\nonumber\\
 &&-\frac{tn}{2i} \sqrt{A} \sum_{{\bf p}}
\biggl\{\tilde{m}_{y}((P-Q)\hat{x})-\tilde{m}_{y}((Q-P)\hat{x})\biggr\}.
\end{eqnarray}
We first need to minimize this functional to determine the ground state.
Since there is no term linear in $m_z$ it follows that $m_z(q) \equiv 0$
in the ground state.  For ground state calculations we could work
with the $XY$ model in both commensurate and incommensurate cases.
Minimizing with respect to $\tilde m_y$ we find that in the ground
state $\tilde m_y$ depends only on $x$ and that its $x$-dependence is
determined in Fourier space by solving
\begin{equation}
\rho_{s}q_{x}^{2} \tilde m_{y}^{\circ}(q_{x}) + \frac{tn}{2}
\left(\tilde m_{y}^{\circ}(q_{x} + Q_{s}) + m_{y}^{\circ}(q_{x} - Q_{s})\right)
= \frac{t\sqrt{A}}{4\pi\ell^{2}i} (\delta_{q_{x},Q_{s}} -
\delta_{q_{x},-Q_{s}}).
\label{eq:sollat}
\end{equation}
We restrict our attention here to results which are valid to
leading order in $t$ so that the second term on the left-hand-side
of Eq.~(\ref{eq:sollat}) can be dropped.  This gives
\begin{equation}
m_{y}^{\circ}(q_{x}) =
\frac{t n\sqrt{A}}{2i\rho_{s}Q_{s}^{2}} (\delta_{q_{x},Q_{s}} -
\delta_{q_{x},-Q_{s}})
\label{eq:sollatfo}
\end{equation}
which gives a ground state energy per unit area equal to
\begin{equation}
 \frac{E}{A}=  - \frac{1}{4} \frac{(tn)^{2}}{\rho_{s}(Q-P)^{2}} +
\frac{\rho_{s}P^{2}}{2}
\end{equation}
Minimizing this energy with respect to $P$ we find that the
overall ground state occurs for
\begin{equation}
\frac{P}{Q} = \frac{1}{2} \frac{(tn)^{2}}{\rho_{s}Q^{4}}=
\frac{\pi^4}{512} \left(\frac{Q_{c}}{Q}\right)^{4}
\end{equation}
This result agrees with the analytic expression given in 
Ref.[\onlinecite{bak}] after making several corrections 
of some typos in Eq.~(2.7) and (2.8). 
For large parallel fields $P$ approaches $0$, $Q_{s}$ approaches
$Q$ and the ground state of the system asymptotically
approaches the ground state in the absence of interlayer tunneling.
We emphasize that as $Q$ approaches $Q_c$ from above $\tilde m_y^{\circ}$
can become large and the linearization approximations will fail even in
the ground state.

Having identified the ground state pseudospin functional we are
able to calculate the collective mode energies and response functions.
To leading order in $t$ we find that
\begin{eqnarray}
\dot{\tilde{m}}_{y}({\bf q}) &=& -\frac{4\pi\ell^2 }{\hbar } \frac{\partial
E}{\partial m_{z}(-{\bf q})}\nonumber\\
 &=& -\frac{2}{\hbar n} \left[
-h_{z}({\bf q}) + \left(2\beta + \rho_s q^{2} - \rho_s P^{2}\right)
m_{z}({\bf q}) + \frac{tn}{2} \left[ m_{z}({\bf q} + Q_{s}\hat{x}) +
m_{z}({\bf q} - Q_{s}\hat{x})\right]\right]\\
\dot{\tilde{m}}_{z}({\bf q}) &=& +\frac{4\pi\ell^2}{\hbar } \frac{\partial
E}{\partial\tilde{m}_{y}(-{\bf q})}\nonumber\\
 &=& \frac{2}{\hbar n}
\left[\rho_s q^{2} \delta\tilde{m}_{y}({\bf q}) + \frac{tn}{2}
\left[\delta\tilde{m}_{y}({\bf q} + Q_{s}\hat{x}) +
\delta\tilde{m}_{y}({\bf q} - Q_{s}\hat{x})\right]\right]
\end{eqnarray}
The soliton lattice acts like an internal grating\cite{cotesoliton}
which couples collective excitations of the ground state in the
absence of interlayer tunneling whose wavevectors are separated
by multiples of $Q_s \approx Q$.  The collective modes of the system for
small $t$ will consist of the zone-folded modes of the $t=0$ system
with small corrections due to mode-mode coupling.
The response functions of the system are readily evaluated numerically
from the above equations.  Artificial external gratings are
often\cite{grating} used to allow the infrared light to couple to
finite-wavevector excitations of two-dimensional electron systems.
The soliton lattice appears to offer an opportunity to couple
to the Goldstone collective mode of the $t=0$ system at a wavevector
which can be tuned by the application of an in-plane magnetic field.
We reemphasize that as $Q$ approaches $Q_c$ from above $\tilde m_y^{\circ}$
can become large and the linearization approximations will fail even in
the ground state.

\section{Exact Diagonalization Studies}

\label{sec:exact_diag}

We now turn to a discussion of
exact numerical studies of the commensurate-incommensurate
transition, and in particular the $B_\parallel$ dependence of the energy gap.
The importance of long length scales in this transition
will limit the power of the exact diagonalization approach,
but we are still able to obtain some useful insights into
the experimental results
obtained by Murphy et al.\cite{murphyPRL}.
We simulate a bilayer 2D quantum Hall system in the presence
of a tilted magnetic
field with a finite number of electrons, and perform our
calculations on the torus.
We make use of the formalism developed by Haldane\cite{Ha:85} for the torus
to block-diagonalize the Hamiltonian in each momentum sector.
As noted in Sec.~\ref{sec:parallelB}, the in-plane component of the
magnetic field enters only in the tunneling matrix elements.
In order to keep the momentum a good quantum number,
we are constrained to use only discrete values of the
magnetic field, namely the values which correspond to an integer number of flux
quanta enclosed between the layers.
We therefore study the transition for a fixed value
of $B_\parallel$, and continuously vary the tunneling between the layers.
This is to be contrasted with
the experimental situation in which the tunneling is fixed and
$B_\parallel $ varies continuously.

Fig.~(\ref{fig:mag10_1}) shows
the pseudo-magnetization calculated as a function of tunneling
amplitude for various fixed values of $B_\parallel$,
for ten electrons and $d/\ell=1$.
For fixed $B_{\parallel}$ the commensurate state which optimizes
the tunneling energy at the cost of exchange energy occurs at
large $t$.  We see in Fig.~(\ref{fig:mag10_1}) that the
component of pseudo-magnetization aligned with the effective
Zeeman field increases with increasing
$t$ and decreases with $B_{\parallel}$, even in the commensurate state.
This quantum fluctuation effect is not captured in the classical 
field theoretic results we have presented since,
, strictly speaking, they 
apply only when the tunneling amplitude is small.  The
increase in the effective magnitude of the ordered
moment with tunneling in the commensurate state may
explain the discrepancy between the $t^{1/2}$ behavior
of the ordered moment predicted here and
the approximately linear behavior seen experimentally.\cite{murphyPRL}

In our finite-size studies the
phase transition to the incommensurate state
appears as a level crossing between
states with different Haldane pseudomomenta which is accompanied
by a large decrease in the component of the pseudospin
aligned with the Zeeman field.  In the thermodynamic limit
and close to the phase transition,
the pseudospin in the incommensurate state is expected
to be aligned with the Zeeman field except in the domain walls
of the soliton lattice.  
Thus the spatial average of aligned
moment is expected to decrease continuously at the
phase boundary.

In our finite size calculations the wavevector $Q$ corresponding
to the rotating Zeeman field satisfies,
\begin{equation}
Q \ell = N_{\phi} \biggl({2\pi \over N}\biggr)^{1/2}
\label{eq:qlfs}
\end{equation}
where $N$ is the number of states in a Landau level,
$N_{\phi} = B_{\parallel} d L/\Phi_0$, and $L$ is the edge length
for a square finite size square sample.  For
$N=10$ \, $Q \ell \approx 0.79 $ even for $N_{\phi} =1$.
Even the smallest value of $Q \ell$ that we can consider
is close to the edge of the regime where we expect the
long wavelength theory developed in previous sections to apply.
We {\it do} however still expect the phase transition to occur
outside this regime.  In order to compare these
microscopic calculations more directly with the
long wavelength theory we have evaluated the energy gap as a function
of $t$.  We present our results in terms of an estimate of
the dimensionless parameter $Q \xi$ which plays the central
role in the long-wavelength theory.
\begin{equation}
\label{eq:Qxi}
Q\xi =
\frac{2\pi\,d\ell B_{\parallel}}{\phi_0} \biggl( {2\pi\,\rho_s \over t}
\biggr)^{1/2}.
\end{equation}
To plot our results we have evaluated $Q \xi$ in the Hartree-Fock
approximation; $t = \Delta_{SAS}/2$ corresponding to full pseudospin
polarization and
\begin{equation}
\label{eq:rhohf}
\rho_s^{HF} = -\frac{1}{32\pi^2}\int_0^{\infty}dkV_k\,h(k)k^3.
\end{equation}
where $V_k=(2\pi\,e^2/\epsilon k)\exp(-kd)$, and $h(k)=-\exp(-k^2/2)$.
Both $\rho_s$ and $t$ are substantially
reduced by quantum fluctuations.  Previous estimates from
finite-size exact diagonalization calculations\cite{Ilong} suggest that
$t$ is reduced by a larger fraction than $\rho_s$, so that
the Hartree-Fock approximation should increasingly underestimate
$Q \xi$ as the layer separation increases.
According to the P-T model, the transition should occur when $Q\xi=4/\pi$. This
will allow us to compare the HF prediction with the exact diagonalization
result.

Figs.~\ref{fig:gapvsqxi_05}, \ref{fig:gapvsqxi_1}, and \ref{fig:gapvsqxi_2}
show the $Q\xi_{HF}$ dependence of the energy
gap for $d/\ell=0.5,1$ and $2$ respectively.
Note also that the system studied by
Murphy et al. corresponds approximately to $d/\ell = 1.85$.
The cusp in the energy gap seen in these figures results from
the ground state level crossing and is not obviously associated with the
square root singularities in the energy gap predicted by
Read\cite{read2layer} which would be expected to appear only in much larger
systems than we are able to treat.  Nevertheless it does show the
expected behavior of weakening at larger layer separations
where the spin-stiffness and hence the meron-pair string
energy is weakened.
We also notice that Hartree-Fock theory accurately predicts the
critical field for $d/\ell=0.5$, becomes worse at $d/\ell=1$ and
fails badly at $d/\ell=2$.  The direction of the discrepancy is
in the direction anticipated by the above discussion since the value
of $Q\xi_{HF}$ at the transition is smaller than $4 / \pi$.
These results suggest that the mean source of discrepancy
between the experiments of Murphy {\it et al.} and
the P-T theory with Hartree-Fock parameters is quantum
fluctuations which reduce the ordered moment aligned by the
effective magnetic field.  This reduction is responsible for a proportional
reduction in the energy gained by forming the commensurate state
and reduces the parallel field strength at which the transition
occurs.  For parameters appropriate to the system studied by
Murphy {\it et al.} our P-T theory with $\rho_s$ and $t$ given by
Hartree-Fock theory gives $B_\parallel = 1.3$T whereas the experimental value
is $ B_\parallel^* \approx 0.8T$. From Fig.~(\ref{fig:gapvsqxi_2}),
we find $B_\parallel^* \approx 0.6\,T$, in substantially better
agreement with experiment.

\section{Summary}
\label{sec:summary}

We have presented in this paper a theory of the interesting effects
of a weak interlayer tunneling in double
layer quantum Hall systems that spontaneously develop interlayer coherence
in the absence of tunneling. We have discussed the properties of the
ground state, as well as low energy collective
excitations (neutral and charged) of the
system, using both effective field theory and microscopic Hartree-Fock
approaches. In particular, we have identified a novel
commensurate-incommensurate phase transition driven by an in-plane component
of the external magnetic field, which has been observed in recent
experiments. Our theory is in good qualitative and semiquantitative
agreement with experiments.

\section{Acknowledgments}
\label{sec:ack}

It is a pleasure to acknowledge useful conversations with Daniel Arovas,
Greg Boebinger, Nick Bonesteel, Luis Brey, Ren\'e C\^ot\'e, Jim
Eisenstein, Herb Fertig, Matthew Fisher, Duncan Haldane,
Jason Ho, Jun Hu, David Huse, Sheena Murphy, Phil Platzman,
Nick Read, Scott Renn,
Ed Rezayi, Mansour Shayegan, Shivaji Sondhi, Mats Wallin,
and especially Shou-Cheng Zhang, from whom we have also benifited 
from early collaborations.
The work at Indiana University was supported by
NSF DMR94-16906.  The work at the University of Kentucky was supported
by NSF DMR92-02255.  We are pleased to acknowledge the Aspen Center for
Physics where part of this work was performed.

\addcontentsline{toc}{part}{Tables}

\begin{table}
\caption[]{Charged Spin Texture Energies at $\nu_{T}=1$ for Double Layers
Systems with Tunneling. $\tilde{\rho}_{s}\equiv \rho_{s}/(e^{2}/\ell)$
and $\tilde{t}\equiv t/(e^{2}/\ell)$ where
$\rho_{s}$ is the pseudospin stiffness, $t$ is the renormalized tunneling
amplitude, $\ell$ is the magnetic length, $T_{0}=8\rho_{s}/\xi$ is
the soliton string tension and $\xi
=\left(\frac{2\pi\ell^{2}\rho_{s}}{t}\right)^{1/2}$ is the domain wall
width.}
\begin{tabular}{llll}
Regime & $\tilde{t} \leq 4 \times 10^{3}\tilde{\rho}_{s}^{3}$ & $4 \times
10^{3}\tilde{\rho}_{s}^{3} \leq \tilde{t} \leq 10^{-2}/\tilde{\rho}_{s}$
& $10^{-2}/\tilde{\rho}_{s} \leq t$\\ \tableline
Nature of Charged & Meron Pairs & Finite Length & Single
Particle\\
Excitations & & Domain Line Strings & Excitation\\ \tableline
Excitation Size & $\sim \frac{e^{2}}{8\pi\rho_{s}}$ & $\sim
\sqrt{\frac{e^{2}}{4T_{0}}} \propto t^{-1/4}$ & $\ell$\\ \tableline
Excitation Energy & $\sim 2\pi\rho_{s}$ & $\sim \sqrt{e^{2}T_{0}} \propto
t^{1/4}$ & $t$\\
\end{tabular}
\label{table:regimes}
\end{table}

\addcontentsline{toc}{part}{Figure Captions}

\begin{figure}
\caption{Schematic conduction band edge profile for a
double-layer two-dimensional electron gas system.}
\label{fig:fig1}
\end{figure}

\begin{figure}
\caption{Phase diagram for the double layer QHE system (after
Murphy et al.).  Only samples whose parameters
lie below the dashed line exhibit
a quantized Hall plateau and excitation gap.}
\label{fig:qhe/no-qhe}
\end{figure}

\begin{figure}
\caption{A process in double-layer two-dimensional electron
gas systems which encloses flux from the parallel component
of the magnetic field.  The quantum amplitude for such paths
is sensitive to the parallel component of the field.}
\label{fig:fig2}
\end{figure}

\begin{figure}
\caption{Illustration of a meron pair in the presence of tunneling
which confines the region of spin twist to a relatively narrow
domain wall or `string'.  Each end of the string is a vortex carrying
charge $\pm 1/2$.}
\label{fig:meron_string}
\end{figure}

\begin{figure}
\caption{Magnetization as a function of tunneling,
for several integer numbers of flux quanta between the
layers at $N=10$ and $d/\ell =1$.  The curves are labeled
by the integer number of flux quanta produced by the
parallel component of the field in the finite size system.}
\label{fig:mag10_1}
\end{figure}

\begin{figure}
\caption{The energy gap as a function of $Q\xi_{HF}$,
for $B_\parallel$ corresponding to one enclosed flux
quantum, i.e. $N_{\phi}=1$, for $N=8$ and $d/\ell=0.5$.}
\label{fig:gapvsqxi_05}
\end{figure}

\begin{figure}
\caption{The energy gap as a function of $Q\xi_{HF}$,
for $B_\parallel$ corresponding to one enclosed flux
quantum, i.e $N_{\phi}=1$, for $N=8$ and $d/\ell=1$.}
\label{fig:gapvsqxi_1}
\end{figure}

\begin{figure}
\caption{The energy gap as a function of $Q\xi_{HF}$,
for $B_\parallel$ corresponding to one enclosed flux
quantum, i.e $N_{\phi}=1$, for $N=8$ and $d/\ell=2$.}
\label{fig:gapvsqxi_2}
\end{figure}


\begin{references}
\bibitem{smearingarg}  In thinking through the details of this
argument, it may be helpful to the reader to recall that the positions
of the electrons in different guiding center orbits are uncertain
by about one magnetic length.  Thus two guiding centers separated
by a distance of order
$\ell$ can have a finite phase difference and still have the electrons
approach each other closely.
\bibitem{order-of-limits}  Note the order of limits here.  The fact that
the order matters really defines what we mean by spontaneously broken
symmetry.
\bibitem{commentrenorm}  We are considering for the moment the case of zero
temperature so that $B_\parallel$ is unrenormalized by thermal fluctuations.
We also treat the problem classically.
\bibitem{tunneldiff}  The matrix element for tunneling is reduced
by a factor $t\longrightarrow te^{-Q^2\ell^2/4}$ so there is an effect
of parallel fields in the one-body portion of the Hamiltonian.
However this is a small effect even at $B_\parallel^*$.
\bibitem{josephsonnote1} Early discussions on the Josephson like effects
in double layer quantum Hall systems
have been presented by Wen and Zee[\onlinecite{wenandzee}], and Ezawa and
Iwasaki[\onlinecite{ezawa}].
\end{references}
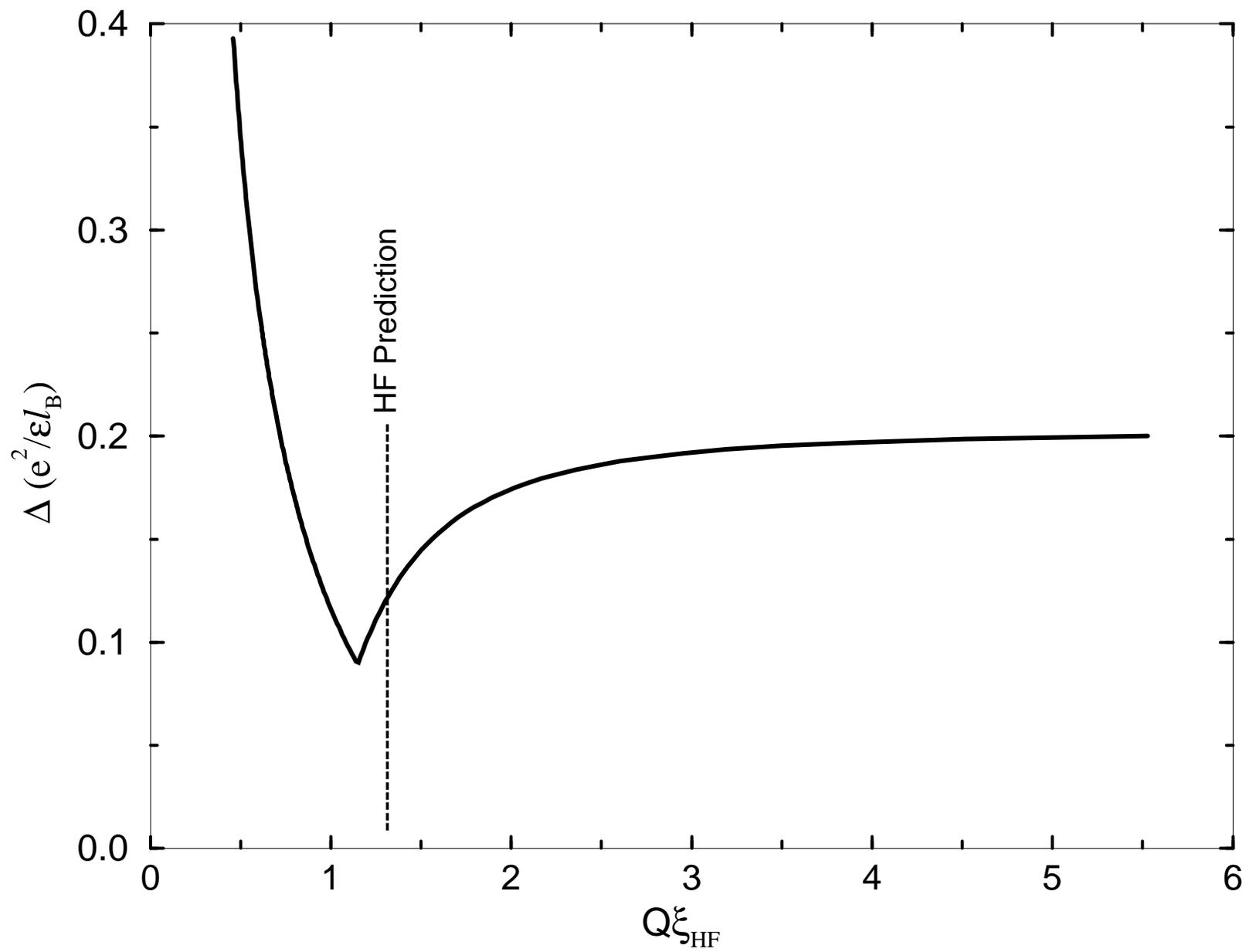
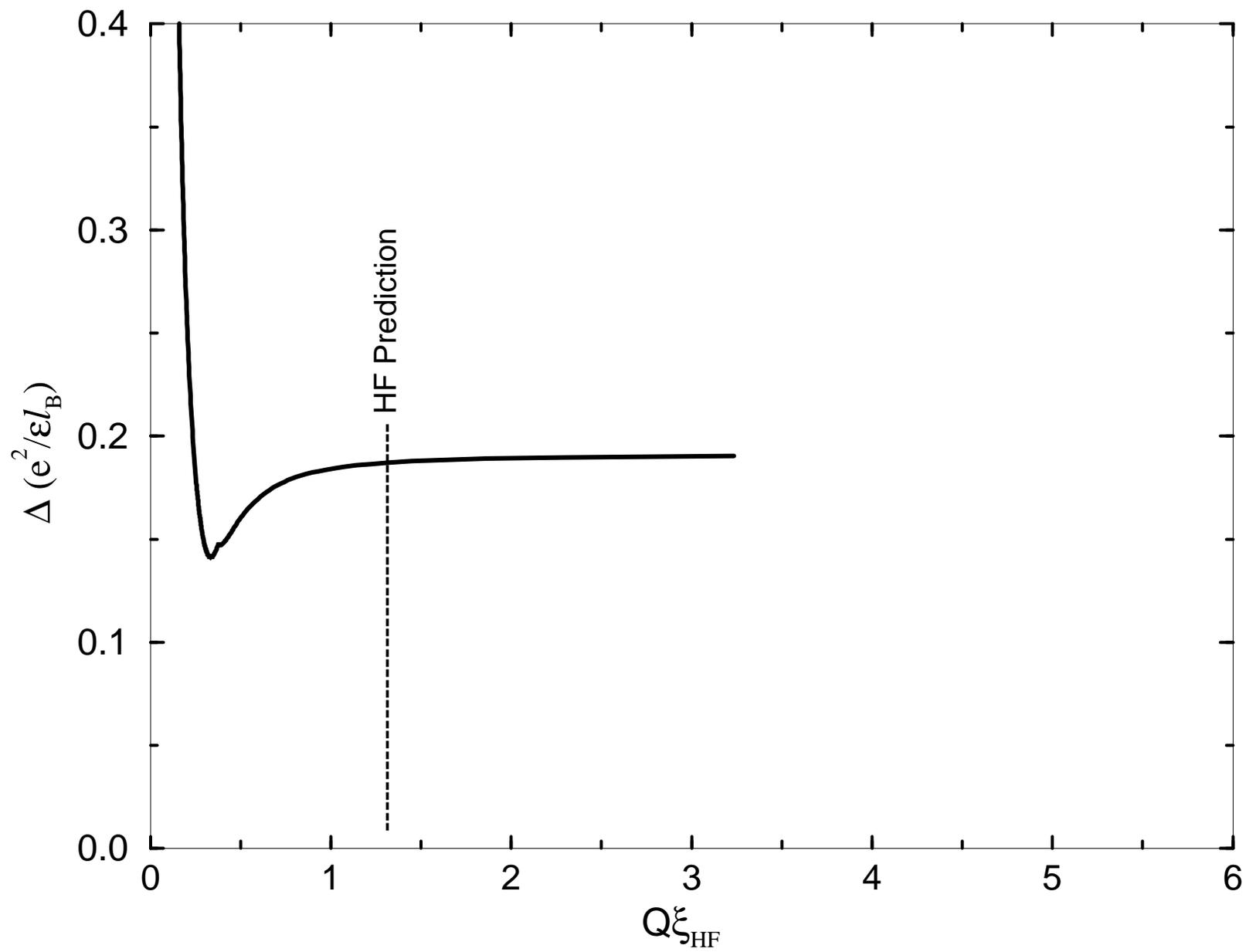
\end{document}